\documentclass[10pt,a4paper,prb,twocolumn,footinbib,floatfix]{revtex4}
\usepackage{graphicx}
\usepackage{amsmath} \usepackage{amssymb}

\usepackage[english]{babel}
\usepackage{times}
\usepackage{epsfig} 
\usepackage{epstopdf} 
\usepackage{color}

\newcommand{\ket}[1]{\left| #1 \right\rangle }  \newcommand{\bracket}[3]{ \left\langle #1 \middle| #2 \middle| #3
\right\rangle }

\begin{document} 

\title{Effect of  hydrogen bonding on infrared absorption intensity}

\author{Bijyalaxmi Athokpam and Sai G. Ramesh}
\affiliation{Department of Inorganic and Physical Chemistry,
Indian Institute of Science, Bangalore 560 012, India}

\author{Ross H. McKenzie}
\email{email: r.mckenzie@uq.edu.au}
\homepage{URL: condensedconcepts.blogspot.com}
\affiliation{School of Mathematics and Physics, University of Queensland,
  Brisbane 4072, Australia} \date{\today}
                   
\begin{abstract}
We consider how the infrared intensity of an O-H stretch in a hydrogen bonded
complex varies as the strength of the H-bond varies from weak to
strong.  We obtain trends for the fundamental and overtone transitions as a
function of donor-acceptor distance $R$, which is a common measure of H-bond
strength.  Our calculations use  a simple two-diabatic state model that permits
symmetric and asymmetric bonds, i.e. where the proton affinity of the donor
and acceptor are equal and unequal, respectively.  The dipole moment function
uses a Mecke form for the free OH dipole moment, associated with the diabatic
states.  The transition dipole moment is calculated using one-dimensional
vibrational eigenstates associated with the H-atom transfer coordinate on the
ground state adiabatic surface of our model. Over 20-fold intensity enhancements for the
fundamental are found for strong H-bonds, where there are significant non-Condon
effects.  The isotope effect on the intensity yields a non-monotonic H/D
intensity ratio as a function of $R$, and is enhanced by the secondary geometric
isotope effect.  The first overtone intensity is found to vary non-monotonically
with H-bond strength; strong enhancements are possible for strong H-bonds.
Modifying the dipole moment through the Mecke parameters is found to have a
stronger effect on the overtone than the fundamental. We compare our findings
with those for specific molecular systems analysed through experiments and
theory in earlier works. Our model results compare favourably for strong and
medium strength symmetric H-bonds.  However, for weak asymmetric bonds we find
much smaller effects than in earlier work.
\end{abstract}

\pacs{}
\maketitle 

\section{Introduction}

A well-known signature of the O-H$\cdots$O hydrogen (H) bond, in addition to the
red-shift of the O-H stretch frequency, is a strong increase in the absorption
intensity
of the infrared band of this mode. \cite{Arunan:2011,Pimental:book}
References
\onlinecite{Marechal:1968, Bratos:1975, Bratos:1991,
Fillaux:1983, Iogansen:1999, Burnham:2006, Howard:2006, Scharge:2008,
Kollipost:2014} are but a subset of the many works that have previously
addressed this effect. 
The work by Iogansen  \cite{Iogansen:1999}
 is particular in that it established an empirical
relation between the  hydrogen bonding energy and the intensity of the
infra-red absorption of the O-H stretching mode for a wide range of compounds:
\begin{equation}
	\Delta H = -12.2 \Delta (A^{1/2}-A_0^{1/2}),
\label{empirical}
\end{equation}
where $\Delta H$ is the enthalpy (kJ/mol) of H-bond formation and $A$ and $A_0$
are the intensities (in units of
$10^4$ cm mmol$^{-1}$=100 km/mol) of O-H stretch in the presence and
absence of the H-bond respectively. This holds for energies varying by a factor
of 200 (between about 0.3 and 60 kJ/mol), thus spanning from weak to strong
H-bonds. Ratajczak, Orville-Thomas, and Rao  \cite{Ratajczak:1976} considered a
theoretical basis for the empirical relation given
in equation (\ref{empirical}) using Mulliken’s
charge transfer theory. Rozenberg  \cite{Rozenberg:2014} recently suggested a relation
between H-bond enthalpy and electron density at the bond-critical point from
atoms-in-molecules theory, and thereby an indirect linear relation between
intensity and electron density. Fillaux \cite{Fillaux:1983}, though primarily concerned
with the theory of H-bond band shapes, 
conjectured a non-monotonic
relationship between intensity $A$ and the donor-acceptor distance $R$, with a maximum
around $R \simeq 2.6$ \AA.

Bratos et al. reviewed experiments describing the variation of
the intensity enhancement with the strength of the H-bond \cite{Bratos:1991}.
For weak H-bonds ($R > 2.8$\AA)
 they find enhancement in the range of about 5 to 10. For
medium strong H-bonds ($R \sim 2.6-2.8$ \AA), it is enhanced by 10 to 15, while for strong
H-bonds
it becomes as large as about 30.  For strong symmetrical H-bonds
($R < 2.6$ \AA), the
enhancement of $A$ decreases by about 10 when $R$ decreases from 2.5 to 2.45
\AA, consistent with a non-monotonic dependence on $R$.
However, estimating the
intensity accurately is difficult due to the broad spectra.

H/D isotope substitution causes a suppression in the O-H stretch intensity. For
free O-H bonds, one anticipates a decrease by a factor of two in the harmonic
picture.  The suppression changes with H-bond strength as well.  For
instance, Bratos et al. \cite{Bratos:1991} state that $A_H/A_D \simeq 2$ for
weak H-bonds, which gets enhanced by $\sim 2.6$ for medium bonds and $\sim 3-5$
for strong H-bonds (see note in Ref.~\onlinecite{Bratos:1991}).

In contrast to the fundamental transition, a number of studies have reported
that the  intensity of the first overtone of the O-H stretch shows a  pronounced
suppression upon H-bonding \cite{Paolo:1972, Scharge:2008, Kollipost:2014,
Howard:2006}.  
Indeed, eighty years ago failure to observe a OH stretching overtone
was correlated with the presence of an H-bond.\cite{Hilbert:1936}
Di Paolo et al. \cite{Paolo:1972} explained this in terms of a
balance between mechanical and electrical anhrmonicity. Suhm and co-workers'
studies of a range of alcohol dimers \cite{Scharge:2008, Kollipost:2014} report
fundamental-to-overtone intensity ratios in the  range of 300 to 1000 for the
H-bonded OH stretches, compared to about 10 for the monomeric OH. For
diols, Howard et al. \cite{Howard:2006} found that the suppression increases for
the donor O-H with H-bond strength from ethane- ($\sim 15$) to propane- ($\sim
83$) to butanediol ($\sim 500$). The acceptor O-H has a smaller value of about 7. 
We parenthetically note that the study of overtones is interesting
in its own right: Heller \cite{Heller:1999} pointed out that overtone excitation
is a purely quantum effect, associated with dynamical tunneling, just like
reflection above a potential barrier.  Lehmann and Smith \cite{Lehmann:1990} and
Medvedev \cite{Medvedev:2012} explicitly showed how the transition probability
for overtone excitation (i.e. the relevant transition matrix element) is
dominated by the semi-classical dynamics in the classically forbidden region of
the potential, particularly the inner wall.

In this paper we study the intensity variation of the O-H stretch transition
with H-bond strength using a simple one-dimensional two-state diabatic model
potential. Section~\ref{sec:theory} discusses the Condon approximation, briefly
describes the diabatic model, the computational details, and gives the form of
dipole moment function.  Section~\ref{sec:results} presents results for the O-H
fundamental intensity variation, isotope effect on the fundamental intensity,
the first overtone intensity variation, and the effect of modifying the dipole
function shape.
In Section \ref{sec:comparison} we give a detailed comparison of our results
with previous theoretical and experimental works. We offer some remarks in the
concluding section.

\section{Computation of the infrared intensity} 
\label{sec:theory}

The intensity of a vibrational transition $j \leftarrow i$ is experimentally
obtained as the integral molar absorption coefficient over the corresponding spectral
band, \cite{Bernath:book, Iogansen:1999} 
\begin{equation} 
	A_{ji} = -\frac{1}{c \, \ell} \int \ln T(\tilde{\nu}) d \tilde{\nu}. 
	\label{eq:Adefexp} 
\end{equation} 
where $T$ is
the transmittance, $c$ in the concentration, and $\ell$ is the path length. The final
unit for $A_{ji}$ is km/mol. Time-dependent perturbation theory yields the
theoretical expression for the intensity as \cite{Bernath:book} 
\begin{equation}
A_{ji} = \frac{2 \pi^2}{3 \epsilon_0 h c}\tilde{\nu}_{ji} |\mu_{ji}|^2,
\label{eq:Adefth} 
\end{equation} 
where the transition dipole matrix element
\begin{equation} 
	\mu_{ji} = \int \ dr \phi_j^*(r) \mu_g (r) \phi_i(r) 
	\label{matrix} 
\end{equation}
where $\nu_{ji}=E_j - E_i$ and $ \phi_i(r) $ is a vibrational wave function and
$r$ denotes all the nuclear co-ordinates.  For notational simplicity we suppress
the vector character of the dipole moment.  Here $\tilde{\nu}_{ji}$ is in
cm$^{-1}$, $\mu$ is in Debye, and the final units of $A_{ji}$ are again km/mol. 

In order to have a sense of the magnitude of $A$, we note that simple alcohol
monomers are reported to have experimental and theoretical gas phase fundamental
intensities in the range of about 25 km/mol \cite{Zou:2014, Phillips:1998,
Provencal:2000}. The corresponding gas phase dimers show an intensity
enhancement of about an order of magnitude \cite{Howard:2006, Scharge:2008,
Provencal:2000}. Experiments by Kuyanov-Prozument et al. on water dimers gives
values of 44 and 144 km/mol for the monomer (asymmetric stretch) and dimer,
respectively \cite{Kuyanov-Prozument:2010}.

The {\it Condon approximation} \cite{Condon} is often applied to
Eq.~\eqref{matrix}. The dipole function enters the intensity expression through
its first derivative alone:
\begin{equation}
	 \mu_{ji} \simeq
	 \mu^C_{ji}
	 \equiv \frac{\partial\vec{\mu_g}(r_{eq}) }{\partial r} r_{ji},
	 \label{eq:condon1}
 \end{equation}
where $r_{eq}$ is the equilibrium O-H bond length (i.e. the
value of $r$ at which the potential energy is a minimum along the O-H stretch).
Deviations from the Condon approximation are also known as electrical anharmonicity.

The Condon approximation leads to several further analyses. 
(1)
There are two distinct physical mechanisms whereby H-bonding can increase the
intensity. The first is by increasing the dipole derivative.  The second is by
increasing the position matrix element, which will be related to the amount of
zero-point motion.
(2) 
If the nuclear wave functions are harmonic, then the only vibrational transition
with non-zero intensity is that of the fundamental (i.e. from the ground state
$i=0$ to the first vibrational excited state, $i=1$). There are no overtones,
i.e. higher harmonics. This is known as the \emph{double harmonic approximation}.
(The first is the Condon approximation).  In reality, all potential energy
surfaces are anharmonic and so this leads to the presence of weak
overtones in IR spectra. Their intensity can be used to estimate the amount of
anharmonicity, both in the potential and the dipole moment surface (i.e.
deviations from Condon).  In the harmonic approximation, $|r_{01}|^2 \sim
\hbar/(m \omega) \propto 1/\sqrt{m}$, where $\omega$ is the harmonic frequency of
the oscillator.  This gives  a limiting value for the isotope effect on the
fundamental intensity: $A_H/A_D = 2$.
(3) 
The Thomas-Reiche-Kuhn (TRK) sum rule \cite{Wang:1999} relates the oscillator
strengths of the ground-to-excited-state transitions: 
\begin{equation} 
	\sum_j (E_j-E_0) |r_{j0}|^2 = \frac{\hbar^2}{2m}. 
	\label{eq:trksum} 
\end{equation} 
$E_j$ is the energy of the $j^{\rm th}$ vibrational state and $m$ is the reduced
mass of the oscillator.  This is true for any potential.  In the Condon
approximation (Eq.~\eqref{eq:condon1}), the terms in the summation differ from
the intensity (Eq.~\ref{eq:Adefth}) by a common pre-factor the dipole
derivative. 
Generally, the sum will be dominated by the fundamental.
Eq.~\ref{eq:trksum} emphasizes the role of vibrational
(mechanical) anharmonicity in the ratio of fundamental-to-overtone intensities.

The intensities of overtones involve contributions from both electrical and
mechanical anharmonicities. Early work by di Paolo et al.  \cite{Paolo:1972}
showed that, for a Morse oscillator with second-order dipole expansions, the two
anharmonicities have cancelling influences for the first overtone's intensity
while being additive for the fundamental. Ref.  \onlinecite{Scharge:2008} found
that the relative signs of the dipole moment first and second derivatives for
H-bonded OH of 2,2,2-trifluoroethanol dimer to be in agreement with this notion. 

Recent works have quantified the effect of the two anharmonicities on the
fundamental and overtone intensities of infrared lines for simple molecules. 
For example,  Vazquez and Stanton \cite{Vazquez:2006} studied
H$_2$O and HFCO, while Banik and Prasad \cite{Banik:2012} studied H$_2$O and
H$_2$CO.  For these simple isolated molecules the effect of the anharmonicities
on the intensity of the fundamental is typically only a few per cent.

Whether the assumption of slow variation of the dipole moment over the relevant
length scale of the oscillator wave functions is applicable for H-bonded
complexes, at various H-bond strengths, is a relevant question. To the extent
that it is valid, the other contribution to the intensity is the mechanical
anharmonicity. This increases as H-bonding strengthens, which results in an
increase in intensity as well.  However, there are significant cases of
non-Condon  effects.  Schmidt, Corcelli, and Skinner \cite{Schmidt:2005} found
that for the OH stretch in liquid water one needs to take into account the
dependence of the dipole moment on the nuclear co-ordinates of the surrounding
water molecules.

\subsection{Diabatic state model for H-bonding}
\label{sec:diabmodel}

In this work, we use the two-state diabatic state model for 
linear symmetric O-H$\cdots$O
H-bonds from recent work by McKenzie \cite{McKenzie:2012}. It was shown in subsequent
work \cite{McKenzie:2014, McKenzie:2015} that it affords a quantitative description
of the correlations observed \cite{Gilli:book} 
between the OO distance ($R$) and OH bond lengths ($r$), the frequencies
of OH vibrations (both stretch and bend), 
and H/D isotope effects for a diverse range of chemical 
compounds \cite{McKenzie:2014, McKenzie:2015}. 
We use the same notation and parameters as in Ref. \onlinecite{McKenzie:2014}.

For a O-H$\cdots$O complex, 
the Hamiltonian with respect to the diabatic states,
$\left|\text{O-H}\cdots\text{O}\right\rangle$ and 
$\left|\text{O}\cdots\text{H-O}\right\rangle$,
is given as 
\begin{equation} 
	H = \begin{pmatrix} V(r) & \Delta(R) \\ \Delta(R) & V(R-r) + V_o \end{pmatrix}
\label{eq:diabham} 
\end{equation} 
The coordinates $r$
and $R$ are the OH and OO distances, respectively, and $r_0$ is the equilibrium
free OH distance of 0.96 \AA. $V(r)$ is Morse potential with a depth ($D$) of
120 kcal/mol, an exponential parameter ($a$) of 2.2 \AA$^{-1}$, corresponding to
a harmonic frequency of 3600 cm$^{-1}$. Its arguments $r$ and $R-r$ in
Eq.~\eqref{eq:adpot} point to the O-H$\cdots$O and O$\cdots$H-O diabats,
respectively. 
$V_o$ is a vertical offset. In this work, we consider both
symmetric and asymmetric cases; more details are at the end of this subsection.
The coupling between the diabats is
given as $\Delta(R)=\Delta_1 \exp(-b(R-R_1)$, with $\Delta_1=48$ kcal/mol,
$b=a$, and $R_1=2r_0+1/a \approx 2.37$ \AA. We note that this is the abbreviated
form of the coupling: The full form contains an angular dependence on the two
HOO angles as well \cite{McKenzie:2012}.

We treat the donor-acceptor distance  $R$ as a control parameter. The electronic
ground state for the above Hamiltonian is given as 
\begin{equation} 
\ket{\Psi_g(r|R)} = 
- \sin \theta(r|R)
\left|\text{O-H}\cdots\text{O}\right\rangle 
+ \cos\theta(r|R)
\left|\text{O}\cdots\text{H-O}\right\rangle
\label{eq:Psig}
\end{equation} 
where the angle is given by
\begin{equation}
\tan 2\theta(r|R) = \frac{2\Delta(R)}{V(r) - V(R-r) - V_o}.
\label{eq:thetadef} 
\end{equation} 
We note that this form for the ground state of the 
electronic wavefunction allows for the charge transfer character
of a H-bond, as emphasized by Thompson and Hynes \cite{Thompson:2000}.
The potential curve corresponding to this state is
\begin{multline} 
\epsilon_-(r,R) = \tfrac{1}{2}\left[V(r) + V(R-r) + V_o\right] \\
 - \tfrac{1}{2}\left[(V(r) - V(R-r) - V_o)^2 + 4\Delta(R)^2 \right]^{\frac{1}{2}}.
\label{eq:adpot}
\end{multline}
For $V_o=0$, this yields a symmetric double well. This is a suitable choice for
strong bonds, since the H atom is essentially shared by the donor and acceptor.
In other words, the respective $pK_a$'s are about the same \cite{Gilli:book}.
However, for weak H-bonds, a sizeable $V_o$ is more appropriate. In this work,
we consider $V_o = 0$ at all $R$, and $V_o = 50$ kcal/mol for $R
\geq 2.7$ \AA. In the latter case, we discuss the variability of the results
with asymmetry.

\subsection{Vibrational eigenstates}
\label{sec:vibstates}
The vibrational eigenstates used in this work to compute infrared intensities
are the 1-D vibrational eigensolutions for a H/D atom on $\epsilon_-(r|R)$.
They are calculated using sinc-DVR functions. For the $V_o=0$ case, the
potential is a symmetric double-well.
Hence, the solutions are labelled $\phi_{n\pm}$ or $n^{\pm}$, where
$\pm$ indicates symmetric and antisymmetric tunnel-split doublets. Of course,
such a label is truly relevant only if the energy levels are well-below the
barrier height. However, we use these labels at all $R$; see
Ref.~\onlinecite{McKenzie:2014} for further details. 
For the asymmetric cases, we simply drop the $\pm$ subscript.

Of primary interest in this work are the ground ($\phi_{0\pm}$ or $\phi_0$),
first excited ($\phi_{1\pm}$ or $\phi_1$), and second excited ($\phi_{2\pm}$ or
$\phi_2$) states.
Transitions between these states define the fundamentals and
overtones we analyse.

When H is replaced with D, a secondary geometric isotope effect (SGIE) is
observed, wherein the O-O distance changes
\cite{Robertson:1939,Ichikawa:2000,Sokolov:1988}. This is purely  a quantum
effect based on the vibrational zero-point energy gradients. Within our
diabatic model, as the H-bond strengthens from $R=3.0$ {\AA} to about $R=2.45$ \AA,
deuteration leads to a progressive increase in the O-O equilibrium distance of
up to about 0.04 \AA. Though small in magnitude, it was found to yield
significant H/D frequency effects \cite{McKenzie:2014}. This is because
changing $R$ changes the shape of the OH stretch potential, and small changes
in $R$ are particularly significant in the regime of 
low-barrier H-bonds where the energy barrier is comparable to
the OH stretch zero point energy.
For $R\lesssim2.4$ \AA,
the direction of the trend is found to be reversed. In analysing the role of
SGIE on the intensities, the eigenenergies and wavefunctions for deuterium are
computed at two distances, to wit, without and with the model-estimated O-O
distance change. This is carried out only for the symmetric case, $V_o=0$.

\subsection{Dipole moment for an H-bond, Condon approximation}
\label{sec:dipole}

For the two diabats, the O-H dipole moments point in opposite directions.
For a symmetric H-bond, it is then evident that the ground adiabatic
state dipole moment function, $\mu_g(r|R)$, would be antisymmetric. To generate such
a dipole function, we assume the following form of the \emph{diabatic} dipole
function:
\begin{equation}
\hat{\mu}_d = \begin{pmatrix} \mu_0(r) & 0 \\ 0 & -\mu_0(R-r) \end{pmatrix},
\label{eq:mudiab} 
\end{equation}
where $\mu_0$ is a
suitable, common form for the dipole moment of both diabats, and the explicit sign
indicates the direction. This is the Mulliken-Hush approach \cite{Nitzan:book} where
there is no cross term in the diabatic representation of $\mu$.
We assume that
the choice of common form of $\mu_o$ for both diabats holds for asymmetric
potentials as well.
This leads to the
definition of adiabatic $\mu_g$ as 
\begin{equation}
\begin{split}
\mu_g(r|R) & = \bracket{\Psi_g}{\hat{\mu}_d}{\Psi_g} \\
		   & = \sin^2 \theta(r|R) \mu_0(r) - \cos^2\theta(r|R) \mu_0(R-r) \\
		   & = \mu_0(r) - \cos^2 \theta(r|R) \left\{ \mu_0(r) + \mu_0(R-r) \right\},
\end{split}
\label{eq:mugdef}
\end{equation}
where from (\ref{eq:thetadef})
\begin{equation}
2\cos^2\theta(r|R) = 1 + \frac{V(r)-V(R-r)-V_o}{\sqrt{[V(r)-V(R-r)-V_o]^2+4\Delta^2}}.
\label{eq:cos2th}
\end{equation}
It remains to choose a form for $\mu_0$.

A simple analytical form of a bond dipole moment function
is that due to Mecke \cite{Mecke:1950}:
\begin{equation} 
	\mu_0(r) = \mu^* r^m \exp(-r/r^*).
	\label{eq:mecke}
\end{equation}
This has the desired limits that it vanishes for small and large $r$. We use the
Lawton and Child \cite{Lawton:1980} parameter values of $m=1$, $\mu^* = 7.85$
D/\AA, and $r^*=0.6$ \AA, originally given for the OH bond in water. The dipole
moment has a negative slope at the equilibrium bond length, $r_0=0.96$ \AA. To a
good approximation, for $r \sim 0.8 - 1.8 $ \AA, which spans the full range of
H-bonds, this dipole moment function is linear \cite{Jakubetz:1989}; compare
Figure 5.14 in Ref. \onlinecite{Grossmann:2008}.  Expanding to first order about
$r_0=0.96$ \AA, we get
\begin{equation}
	\mu_0(r) = \mu_1 - \mu' (r -r_0).
	\label{eq:meckelinear} 
\end{equation}
where $ \mu_1= 1.52$ D
and $ \mu' = 0.95$ D/\AA. 
Note that this linear form corresponds to a Condon approximation for an isolated
OH bond. Although all the results we present in the subsequent sections are with
the full form of Eq.~\eqref{eq:mecke}, we note that using the linearized form of
$\mu_0$ [Eq.~\eqref{eq:meckelinear}] in $\mu_g$ yields dipole functions that are
slightly different (under about 5\%) at various $R$.

The Condon approximation (Eq.~\eqref{eq:condon1}) for $\mu_g(r)$
involves the evaluation of its derivative at $r_{eq}(R)$, which is the
minimum of the adiabatic potential $\epsilon_-(r|R)$ at different $R$. The
approximation would be valid to the extent that this shape of $\mu_g$ is
approximately linear in a sufficiently wide interval about $r_{eq}$. 
Below, we will compare the dipole moment function
$\mu_g(r)$ with the wavefunction shapes at different $R$ to determine if this
is so.

Lastly, we note the selection rules for the fundamental and overtone transitions.
Since $\mu_g$ is antisymmetric in $r$ for all $R$, the allowed transitions involve a
change in the symmetry of the vibrational wavefunction,
i.e., a change in parity.
 We focus on three transitions:
$1^+\leftarrow0^-$, $1^-\leftarrow0^+$,
and $2^+\leftarrow0^-$.
In the next section we discuss the possible identification
of these transitions with the fundamental and first overtone. 
ranges;
See also the discussion in Sec.~III.V of Ref.~\onlinecite{McKenzie:2014}.

\section{Results}
\label{sec:results}

\subsection{Frequency vs H-bond length (R)}

\label{sec:freq}
\begin{figure}[!tbh]
	\includegraphics{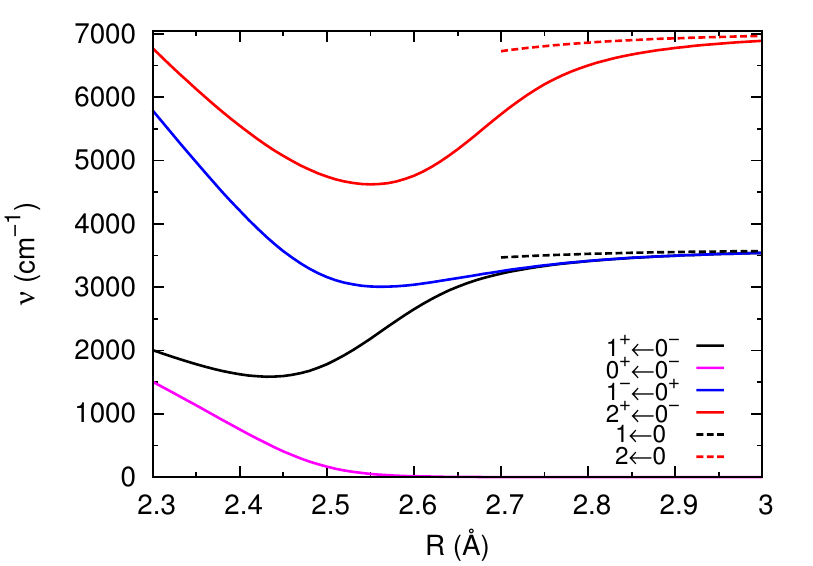}
	\caption{
	Variation of different OH stretch transition frequencies with the donor
	acceptor $R$ for symmetric and asymmetric H-bonds.
	\emph{Solid lines, symmetric H-bonds:} The curves plotted are
	$1^{+} \leftarrow 0^{-}$ (black),
	$1^{-} \leftarrow 0^{+}$ (blue), 
	$2^{+}\leftarrow0^{-}$ (red), and
	$0^{+} \leftarrow 0^{-}$ (magenta).
	For weak bonds ($R > 2.7$ \AA) the black and red curves can be identified
	with fundamental and first overtone transitions respectively. The blue curve
	separates from the fundamental curve (solid black) only when the tunnel
	splitting becomes significant. For moderate bond strengths, the solid black
	curve has the lowest frequency in the experimentally relevant range ($>500$
	cm$^{-1}$) and so is identified as the fundamental. For strong bonds, there
	are large anharmonic effects and the nomenclature of fundamental and first
	overtone is not particularly meaningful.
	\emph{Dashed lines, asymmetric H-bonds:} The $1 \leftarrow 0$ and $2
	\leftarrow 0$ transition frequencies are plotted for $R \geq 2.7 $ {\AA} for
	an asymmetry ($V_o$) of 50 kcal/mol. The effective potential of the lower
	well is a little less anharmonic than for the symmetric case. Consequently,
	the transition frequencies are a little higher.
	}
	\label{fig:omega} 
\end{figure}

We begin with an analysis of the frequencies of different vibrational
transitions as the H-bond strength changes for both symmetric and asymmetric
cases. This is necessary, particular for the symmetric case, to clearly define
what we mean by a \emph{fundamental} and a \emph{first overtone}, since there
are significant anharmonic effects for strong bonds in the symmetric case. For
weak symmetric or weak asymmetric H-bonds, the identification is
straightforward. 

The solid curves in Figure~\ref{fig:omega} are for the \emph{symmetric} case. The
frequency of the $1^{+}\leftarrow0^{-}$ transition frequency is seen to have a
non-monotonic variation with $R$ (black curve).  It is progressively softened
(red-shifted) as the H-bond strength changes from weak ($R\gtrsim2.7$ \AA) to
moderately strong ($R\sim2.5-2.6$ \AA). In the latter region, the barrier height
becomes comparable to the energy of the first few O-H vibrational states, and as
a result the tunnel-splitting is significant. In the very strong H-bond region
($R<2.45$ \AA), the potential becomes roughly square-well like with a very low
or no barrier, and all the vibrational states are energetically well separated.
Hence the $1^{+}\leftarrow0^{-}$ curve turns upward.  For moderate bond
strengths, the black curve has the lowest frequency in the experimentally
relevant range ($> 500 $ cm$^{-1}$) and so is identified as the
\emph{fundamental}.  The above discussion is based on Figure 3 in Ref.
\onlinecite{McKenzie:2014} which shows the different potentials and low-lying
vibrational energies for $R= 2.3, 2.45, 2.5, 2.9$ \AA.

Also shown in Figure~\ref{fig:omega} (top panel) is the $2^+\leftarrow0^-$ transition
frequency (red curve). This, too, has a non-monotonic dependence on $R$.  For
weak bonds, this can be identified as the \emph{first overtone} as it has
roughly twice the frequency of the fundamental.  But it turns upward sooner
compared to the fundamental since the energy of the $2^+$ state, moves higher
than the barrier before the $1^+$ state does.  In the moderate H-bond region,
due to significant tunnel-splitting, the $1^-\leftarrow0^+$ transition frequency
(blue curve) clearly separates from the fundamental curve.  The definition of
the first overtone in this region becomes ambiguous due to the large
anharmonicity of the  potential.  We discuss the intensities for each of these
vibrational transitions in Section~\ref{sec:overtone}. Like the frequencies,
they all have a non-monotonic dependence on $R$.

We also note that for strong bonds with $R \lesssim 2.5$ \AA, the splitting of
the $0^+$ and $0^-$ levels becomes larger than 500 cm$^{-1}$, which is larger
than the thermal energy, $k_B T$ at room temperature. This means that the
population of the $0^-$ level will be reduced by a Boltzmann factor of order
0.1. In an experiment, there will be a corresponding reduction in the measured
IR absorption intensity associated with transitions from this level. In order to
highlight changes in the dipole matrix element, our plots do not take this thermal
effect into account.

For the \emph{asymmetric} case, the chosen $V_o$ value shifts the right diabat
in Eq.~\eqref{eq:diabham} above the energy of the Morse overtone level of the
left (unshifted) diabat. The resulting ground state potential therefore has
\emph{single} and unambiguously identifiable ground, fundamental, and overtone
levels. The corresponding wavefunctions are also largely localized on the left
side.  The fundamental and overtone transition frequencies as a function of $R$
are plotted as red and blue dashed lines in Figure~\ref{fig:omega}. The plots
stop at 2.7 {\AA} since we consider asymmetry only in the weak H-bonding regime.
It is of note that the asymmetric fundamental is higher by 256 and 30 cm$^{-1}$
compared to the symmetric case at 2.7 and 3.0 \AA, respectively. The
corresponding values for the overtone are 996 and 75 cm$^{-1}$. Both sets are
consistently higher. A major part of these differences is due to the lower
harmonic frequency of ground state potential minimum for the $V_o=0$ case than
for $V_o>0$: The diabats are more mixed with decreasing asymmetry and 
at shorter $R$ in general. A smaller role is played by the
effective anharmonicity of the ground state potential well, which reduces
(to simply the anharmonicity of the Morse potential for the diabatic state) with
increasing $V_o$.

\subsection{Intensity of the fundamental transition and Condon breakdown} 
\label{sec:FI}

\begin{figure}[!tbh]
\includegraphics{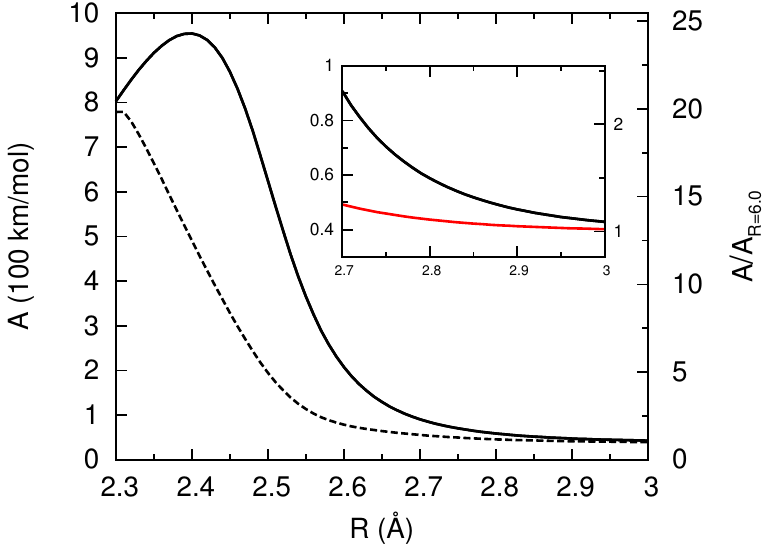}
\caption{
	Intensity of the $1^{+}\leftarrow0^{-}$ transition as a function of
	$R$.  The left axis is intensity in (100 km/mol) units and the right axis is
	the intensity scaled with respect to its value for $R = 6.0$
	{\AA} (non-H-bonded OH).  The solid line is the intensity obtained through
	the full matrix element, Eqn.\eqref{matrix}, while the dashed line is
	obtained using the Condon approximation, eqn.  \eqref{eq:condon1}.  The
	inset is a blow-up of the curve, showing the relatively small intensity
	enhancements in the weak H-bond regime. Also shown in red in
	the inset is the trend when $V_o=50$ kcal/mol.
	}
\label{fig:fund}
\end{figure}

For symmetric H-bonds,
our calculation of the intensity of the $1^{+}\leftarrow0^{-}$ (fundamental)
transition using eqn.~\eqref{eq:Adefth} is shown by the solid line in
Figure~\ref{fig:fund}. The non-H-bonded OH intensity value (computed at $R=6.0$
\AA) is about $39$ km/mol, which compares reasonably with the range of about
$20-60$ km/mol reported for O-H stretches for a range of isolated molecules
\cite{Zou:2014, Phillips:1998, Provencal:2000, Banik:2012}.  The intensity
enhancement relative to this value is a little over 2 in the weak H-bond region (see
inset).  As the curve enters the moderately strong H-bond region ($R\lesssim2.6$
\AA), it shows $\sim 5-10$ fold enhancement, reaching $\sim 20$ for strong
H-bonds ($R \approx 2.4$ \AA).  Broadly, this agrees with experimental results
summarised by Bratos et al. \cite{Bratos:1991}. 

\begin{figure}[!tbh] 
	\includegraphics{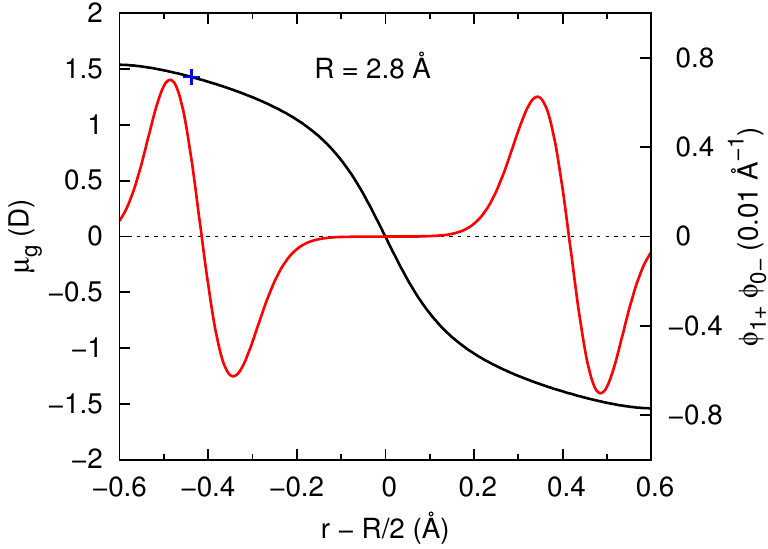}
	\includegraphics{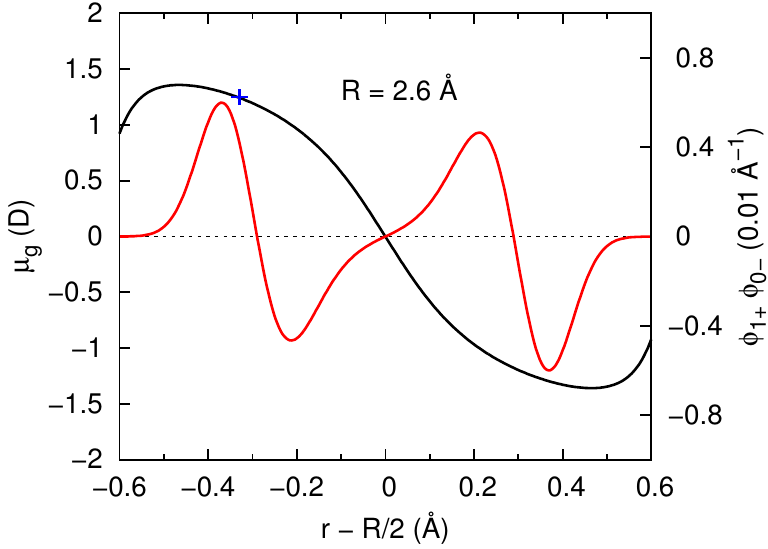} 
	\includegraphics{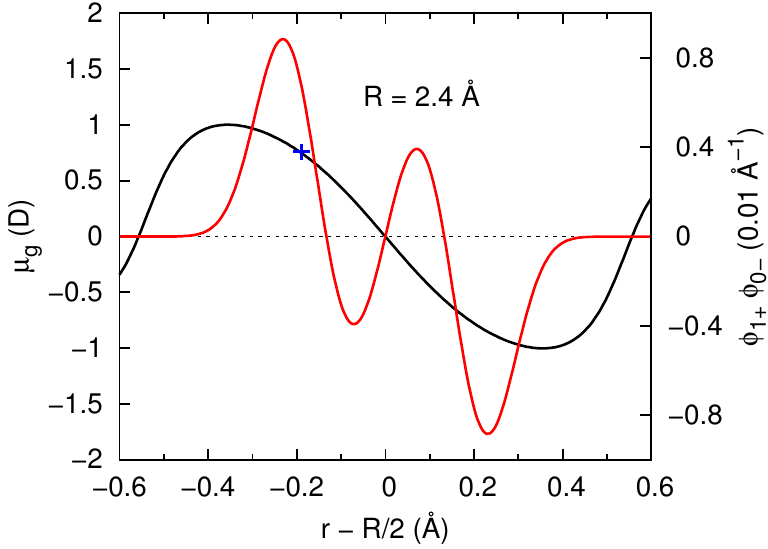}
	\caption{
	Origin of breakdown of the Condon approximation. Plotted are the dipole
	moment function  $\mu_g(r)$ (black line, left axis)	and the wavefunction
	overlap $\phi_{1^{+}}\phi_{0^-}$ (red line, right axis) as a function of $r
	- R/2$ for $R$ = 2.8, 2.6, and 2.4 \AA.  The product of these two functions
	is the integrand in the transition dipole matrix element \eqref{matrix}.  
	As the H-bond strength increases and $R$ decreases, the wavefunction overlap
	has significant weight where the dipole function becomes non-linear.  That
	is, as $r$ gets closer to $R/2$, the slope of the dipole function is
	significantly larger than that at the equilibrium bond length. This
	non-linearity contributes to the enhanced absorption intensity.  The blue
	plus sign marks the classical minimum of the (left-side of the) double-well
	for a given $R$. The dipole derivative in the Condon approximation,
	Eqn.~\eqref{eq:condon1}, is evaluated at this point.
	}
\label{fig:mu_condon} 
\end{figure}

Figure~\ref{fig:mu_condon} shows the contributions to the integrand in
eqn.~\eqref{matrix}, viz. $\mu_g(r)$ and $\phi_{1^+}\phi_{0^-}(r)$, at different
$R$, giving insight into the intensity enhancement with increased H-bond
strength.  These functions are both asymmetric about $r-R/2=0$ at all $R$. Hence
it would suffice to consider only one vertical half of the plots.
The top panel is for $R=2.8$ \AA. Here, $\mu_g(r)$ is mostly linear for a large
O-H distance ($r$) range.  The $\phi_{1^+}\phi_{0^-}$ product function amplitude
is non-zero over roughly the same $r$ range. Its positive and negative regions
have only a small difference in areas, leading to significant cancellations in
the total integral. However, this difference in areas is a little larger than
that at $R=3.0$ \AA, where $\mu_g(r)$ is found to be even more clearly linear in
the relevant $r$ range.  A modest intensity enhancement at $R=2.8$ {\AA}
compared to $R=3.0$ {\AA} is therefore anticipated, and borne out by the plot in
Figure \ref{fig:fund}.

For moderate strength H-bonds ($R \sim 2.6$ \AA, middle panel), $\mu_g$ is seen
to be more non-linear. 
This is a consequence of the shape of the mixing angle $\theta(r)$ with $r$
[compare equations \eqref{eq:Psig}) and \eqref{eq:thetadef}];
with decreasing $R$, it changes less abruptly along $r$ between its diabatic
limits of $0$ and $\pi/2$. As a consequence, the charge transfer character
changes more continuously as the proton moves
from the donor to the acceptor. This is true for $\mu_g(r)$ as well.
Returning to the wavefunction product, the $\phi_{1^+}\phi_{0^-}$ overlap
function has more unequal positive and negative spread at $R=2.6$ \AA.  This
results in less cancellation compared to the case at $R=2.8$ \AA, resulting in
a larger enhancement of intensity.
All these effects becomes stronger still at $R=2.4 $ {\AA} (bottom panel).

The dashed line in Figure \ref{fig:fund} gives the intensity obtained using the
Condon approximation (eqn.\eqref{matrix}). 
The required derivative of $\partial\mu_g/\partial r$ was
evaluated at the classical minimum of the double well for each $R$. In Figure
\ref{fig:mu_condon}, these are marked with blue plus signs.
In the weak H-bond region (large $R$), the intensity calculated through this
approximation is in agreement with the actual value.  But as $R$ decreases the
approximation breaks down and is seen to underestimate the intensity.
Figure \ref{fig:mu_condon} helps explain this Condon breakdown.  For weak
H-bonds, $\mu_g(r)$ is largely linear in the region where $\phi_{1^+}\phi_{0^-}$
has significant amplitude, as seen for $R=2.8$ \AA. Taking a constant dipole
derivative for this case is reasonable. But as the H-bond strengthens,
$\mu_g(r)$ is sufficiently non-linear for $R=2.6$ {\AA}, and even more so at $R
=2.4$ {\AA}.  For these cases, the $\phi_{1^{+}}\phi_{0^{-}}$ overlap curve
becomes less localized, i.e., broader.  This reflects the large zero-point
motion due to the reduced frequency of the OH stretch and the increase in
anharmonicity and tunneling.  Hence, the actual intensity is more enhanced than
that calculated with the Condon approximation.

We now discuss in detail the results for asymmetric H-bonds. The trend for $R$ in the
range
2.7-3.0 {\AA} is shown as the red curve in the inset of Figure~\ref{fig:fund}.
Here, too, there is an enhancement in intensity with decreasing $R$, albeit
smaller than that for the symmetric case. At $R=2.7$ \AA, it is about 1.25 times
that for a free OH. This fraction varies slightly when the asymmetry is changed
to 40 or 75 kcal/mol, the former (latter) leading to higher (lower) intensity.
Insight into why these numbers are all lower than the symmetric case may be
obtained from the work of di Paolo et al.\cite{Paolo:1972} Translating their
notation to ours, the fundamental intensity is proportional to $(\mu'_g -
5b\mu''_g)^2$, where the dipole derivatives are evaluated at the potential
minimum, and $b$ is the (dimensionless) ratio of the cubic anharmonicity to the
harmonic frequency of the well. With $b<0$ being the typical case, and $\mu'_g$
and $\mu''_g$ having the same sign (which is true in our case as well), di Paolo
et {al.} argued that the second term augments the first. Therefore, the potential
and electrical anharmonicity enhance the fundamental intensity. For our case,
the symmetric case has both larger $|b|$ and larger $\mu''_g$ than the
asymmetric one at a given $R$. The underlying cause is the larger mixing of
diabats in the symmetric versus asymmetric models, ultimately leading to the
computed differences in intensities.

\subsection{Isotope effect on the intensity of the $1^{+}\leftarrow0^{-}$
transition} 
\label{sec:isotope}

\begin{figure}[!tbh] 
	\centering 
	\includegraphics{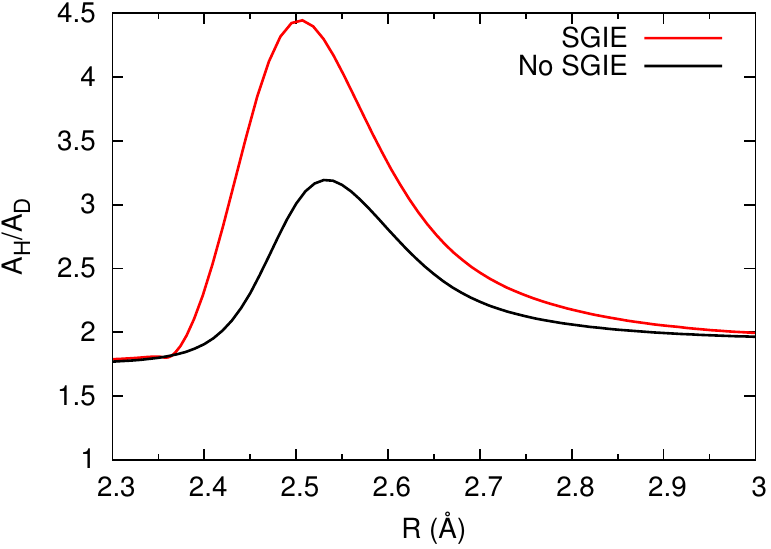}
	\caption{
	Isotope intensity ratio, $A_H/A_D$, for the $1^{+}\leftarrow0^{-}$
	transition  as a function of the donor-acceptor distance $R$. For the red
	(black) curve the secondary geometric isotopic effect (SGIE), i.e, change in
	the donor-acceptor distance $R$, is (is not) taken into account.  The SGIE
	enhances the isotope intensity ratio for low-barrier H-bonds.
	}
\label{fig:ratio}
\end{figure}

Experiments show that the intensity of the fundamental transition of a H-bonded
O-H stretch mode is suppressed upon substituting H by D.\cite{Bratos:1991} The
black curve of Figure \ref{fig:ratio} shows how H/D isotope substitution affects
the intensity of the fundamental,
as calculated for our symmetric H-bond model. (We limit the
analysis to the symmetric case since the effects discussed below are more
important in the medium and strong H-bonds.)
The $A_H/A_D$ ratio shows a non-monotonic dependence on $R$. In the weak H-bond
region, the ratio is almost unaffected as $R$ varies. Also, the Condon
appproximation holds well here: $A_H/A_D = 2$; see Section \ref{sec:theory}. For
H-bonds with moderate strength, the ratio increases reaching a maximum at $R
\simeq 2.53$ \AA. The position of this maximum roughly matches with the minimum
of the frequency ratio in Figure~8 of Ref.~\onlinecite{McKenzie:2014}.  For
still stronger H-bonds, the intensity ratio declines and becomes $\sim1.7$ at
very short $R$. This is attributed to the square-well-like behaviour of the
potential for this range of $R$.  \cite{McKenzie:2014} (For a square-well
potential, the vibrational wavefunctions are independent of mass while the
transition frequencies are mass dependent.  Thus the ratio of intensities will
mainly be due to the frequency ratio, which is approximately 2.)

Another important aspect of the isotope effect is the secondary geometric
isotope effect (SGIE) where the O-O equilibrium distance is changed upon
substituting H by D (Section~\ref{sec:vibstates}). This modifies the adiabatic
potential, which, in turn, also affects the intensity. Therefore, the
experimental quantity that we need to calculate is $A_H(R_H)/A_D(R_D)$, where
$R_D$ is different from $R_H$ due to SGIE.  The red curve of
Figure~\ref{fig:ratio} shows this ratio. Evidently, this ratio is overall larger
compared to the one without SGIE. The maximum is shifted to slightly lower $R$,
and interestingly also roughly corresponds to the H/D frequency ratio minimum
calculated with SGIE in Figure 8 of Ref.~\onlinecite{McKenzie:2015}.  Bratos {et
al.} \cite{Bratos:1991} quotes the $A_H/A_D$ ratio to be about 2, 2.6, and 3-5
for weak, moderate, and strong bonds, respectively. These are in agreement with
our results that include the SGIE.

\begin{figure}[!tbh] 
	\centering 
	\includegraphics{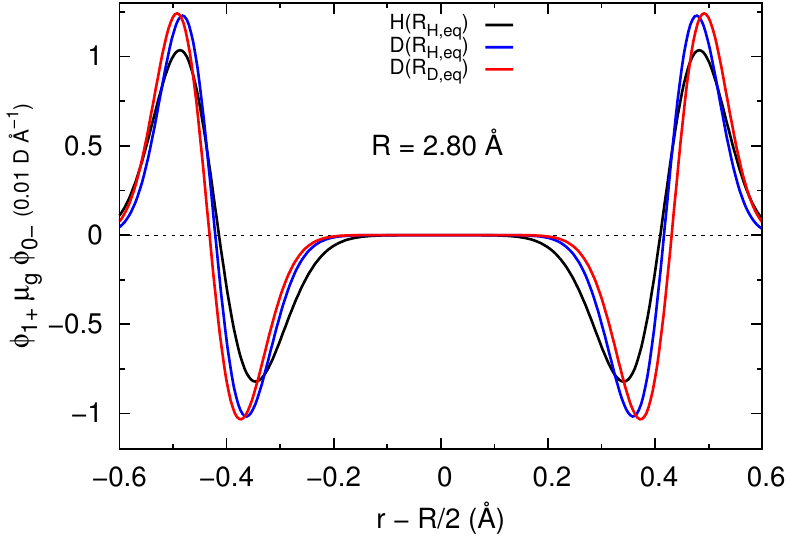}
	\includegraphics{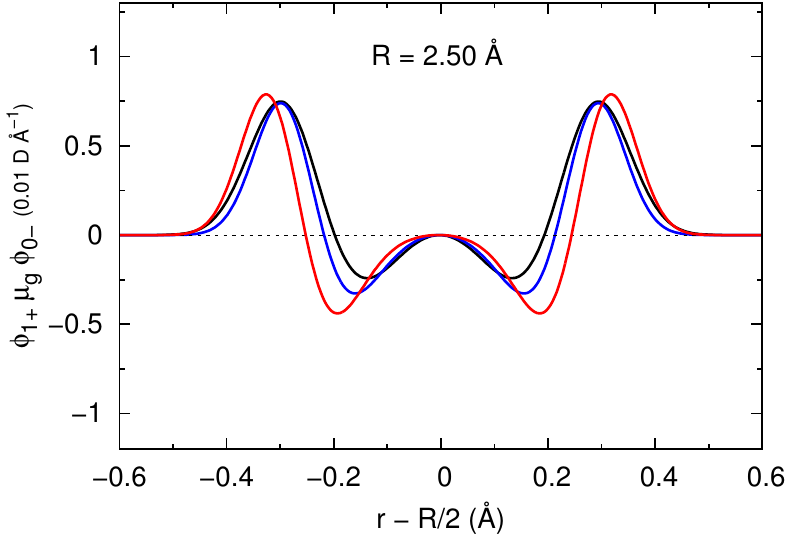}
	\caption{
	Integrand for the transition matrix element $\phi_{1^{+}}\mu_g\phi_{0^{-}}$
	for the H and D isotopes for different $R$.  The black curve is for H
	isotope while red and blue curves are for D isotope with and without the
	secondary geometric isotope effect (SGIE), respectively. The H-bond is
	stronger at lower $R$.
	} 
	\label{fig:mu_sgie} 
\end{figure}

Insight into the observed trend of the $A_H/A_D$ ratio with $R$ can be given by
analysing how the integrand of the transition dipole moment, $\phi_{1^{+}}
\mu_g(r) \phi_{0^{-}}$, varies with $r$ for each isotope at different $R$
values.  This product function is plotted in Figure \ref{fig:mu_sgie} for O-O
distances in the weak ($R=2.8$ \AA) and fairly strong ($R=2.5$ \AA) H-bond
regions. The H (black) and D (blue) curves are without the inclusion of the
SGIE. They are different essentially because H experiences  larger anharmonicity
effects than D. The wavefunctions for H have a greater spread than those for D.
With $\mu_g(r)$ being the same for both, the product function plotted for H in
both panels of Figure \ref{fig:mu_sgie} have larger positive than negative areas
compared to those for D. Therefore, the transition dipole integral is higher for
H than D. On including the SGIE $\mu_{1^+0^-}$ (red curves), one
sees very little change for weak bonding; the integrands with and without this
effect are rather similar.  For strong H-bonds, there is a clear difference. The
resulting  integrals for D are smaller and so $A_H/A_D$ is higher. 

\begin{figure}[!tbh]
	\centering 
	\includegraphics{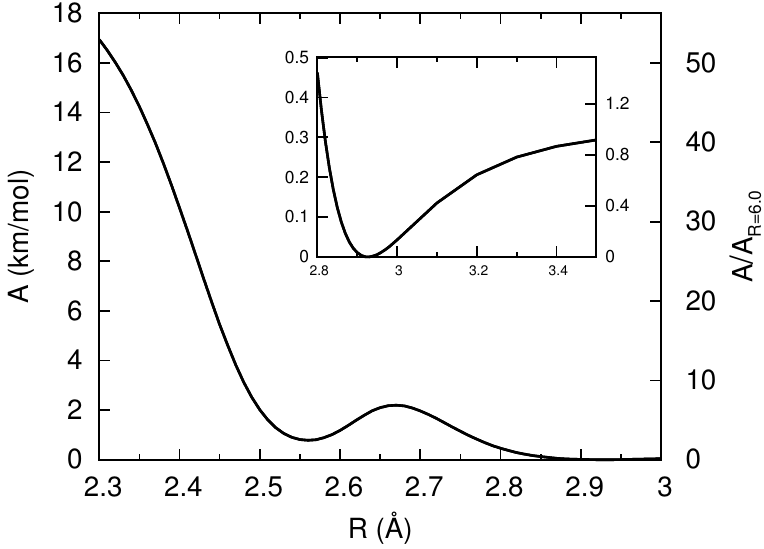} 
	\caption{Dependence of the intensity of the $2^{+}\leftarrow0^{-}$
		transition on $R$.  The left axis is the intensity in units of km/mol
		and the right axis is the intensity scaled by its value
		for $R = 6.0$ {\AA} (absence of H-bond). This clearly shows the
		non-monotonic dependence of the overtone intensity on the strength of
		the H-bond.  Furthermore, for medium to strong bonds, significant
		enhancement of the overtone intensity is possible. The inset shows the
		trend in the weak H-bonding region. Note that there is
		some intensity suppression near and above 3.0 \AA.
	}
	\label{fig:Overtone} 
\end{figure}

\subsection{Overtone intensity} 
\label{sec:overtone}

Figure~\ref{fig:Overtone} shows the intensity of the $2^+\leftarrow0^-$
transition as a function of $R$ for a symmetric H-bond. Its intensity for a
monomeric OH (at $R=6.0$ {\AA} for our model) is about 0.32 km/mol. It has a
complicated non-monotonic dependence on $R$. The inset shows that with
decreasing $R$ the intensity initially drops to zero at about 2.96 \AA, and
thereafter rises rapidly. This initial overtone suppression occurs at a distance
somewhat larger than anticipated based on prior works, which indicate
suppression up to at least 2.8 \AA. We shall see further below that this might
be a consequence of asymmetric H-bonds studied in those works. Continuing to
smaller $R$ or stronger H-bonds, we find the transition intensity going up to
$\sim$ 17 km/mol, which is about a 50-fold enhancement. That the overtone is not
suppressed at all distances, but instead increases to significant values
compared to that for a free OH oscillator, is a new finding in this work. 

\begin{figure}[!tbh]
	\centering 
	\includegraphics{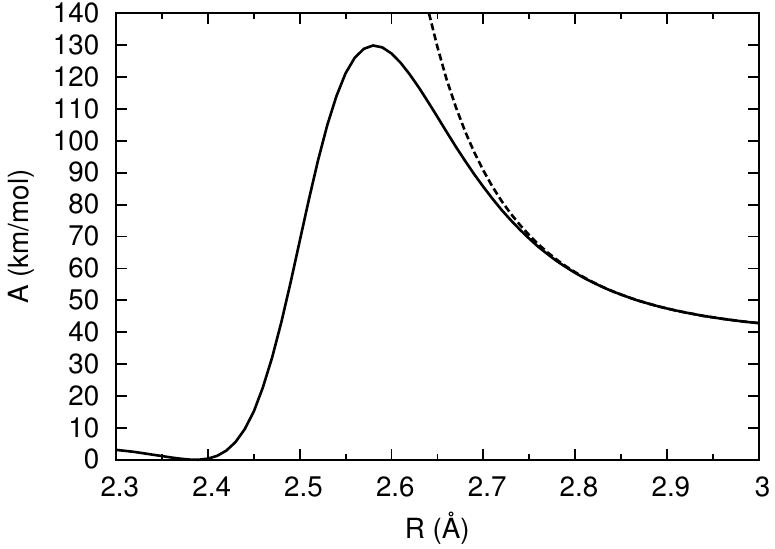}
	\caption{
	Dependence of the intensity of the $1^{-}\leftarrow0^{+}$ transition  on
	$R$.  For this transition, the intensity for weak H-bonds ($ R \sim 2.7$ to
	$3.0$ ~\AA) is the same as that of the fundamental ($1^{+}\leftarrow0^{-}$)
	(shown as the dashed curve) since the tunnel-split ground and excited states
	are hardly distinct (compare Figure \ref{fig:omega}).  It is only for
	stronger H-bonds that this transition may be considered distinct from the
	fundamental.
	} 
	\label{fig:overt2} 
\end{figure}

As argued in Section~\ref{sec:freq}, the $1^- \leftarrow 0^+$ transition may
also be labelled as the overtone for strong H-bonds.  For example, at 2.45
{\AA}, it is this transition that is about twice the fundamental, while the $2^+
\leftarrow 0^-$ transitions has thrice the frequency.  Figure~\ref{fig:overt2}
gives the variation for the intensity of this transition with $R$.  It is of
significance only when $R\lesssim2.6$~\AA, when it becomes distinct from the
fundamental, due to observable tunnel splitting. When this happens the intensity
has a highly non-monotonic  variation with $R$, quite distinct from the
monotonic increase with bond strength of the $2^+\leftarrow0^-$ transition.  In
this region ($R \leq 2.6$ \AA), the $1^- \leftarrow 0^+$ transition has a
generally larger, but rapidly dropping, intensity compared to the $2^+
\leftarrow 0^-$ transition; note the ordinate scale of the two plots.  Thus
observing both frequency (Figure~\ref{fig:omega}) range and intensity
(Figures~\ref{fig:Overtone} and \ref{fig:overt2}) variation will help
distinguish the two overtones.

\begin{figure}[!tbh]
	\includegraphics{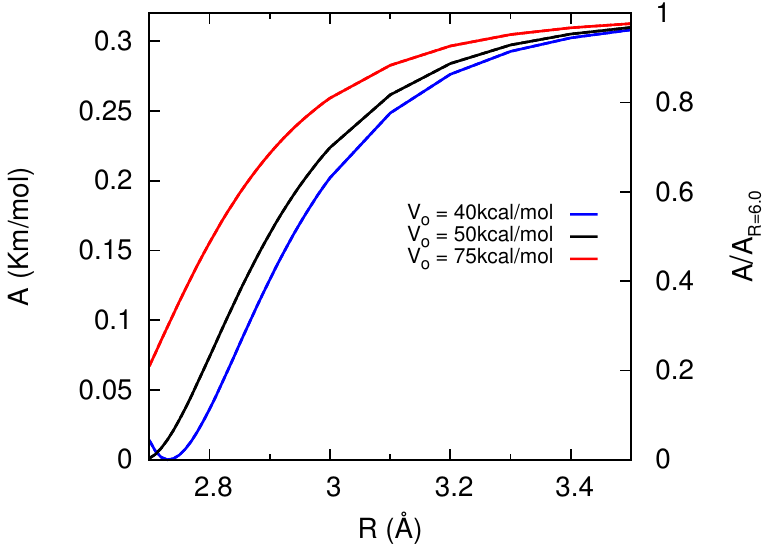}
\caption{	Intensity of the $2^+ \leftarrow 0^-$ overtone transition for different
	asymmetries, {viz.} $V_o=40$, 50, and 75 kcal/mol.  These plots 
	contrast to the symmetric case (inset of Figure \ref{fig:Overtone}), showing
	that the extent and donor-acceptor distance range of overtone suppression in
	an H-bond (relative to a free OH) changes when the double-well potential is
	asymmetric.  The plots also show that these properties can vary with the amount
	of asymmetry ($V_o$).
	}
	\label{fig:overtone_asymm}
\end{figure}

We now return to the weak H-bonding region, and discuss the effect of asymmetry
on the double well potential.  Plots of the $2 \leftarrow 0$ transition using
$V_o = 40$, 50, and 75 kcal/mol are shown in Figure \ref{fig:overtone_asymm}.
Note that the applied $V_o$ are all sizeable compared to the OH dissociation
energy (Morse parameter $D$ here is 120 kcal/mol). Although the shifted right
diabats lie higher than the Morse overtone level (about 25 kcal/mol above the
potential minimum) for all cases, the overtone intensity trends are different
for each $V_o$. Importantly, though, all of them lower the O-O distance range for overtonesuppression to at least 2.8 \AA. It is difficult to ascertain the precise cause of this
change, but our calculations show that overtone properties are rather sensitive
to the shape of the anharmonic potential and the resulting $\mu_g$ as well. Indeed,
it is this sensitivity that leads to the curious trend in Figure
\ref{fig:Overtone}. 

However, a qualitative understanding of the trends between the three $V_o$
values of Figure \ref{fig:overtone_asymm} may be obtained through the work of di
Paolo et al.\cite{Paolo:1972}. They give the overtone intensity to be
proportional to  $(\mu'_gb + \mu''_g)^2$. (See the end of Section \ref{sec:FI}
for the notation.) As such, with $b<0$ and the derivatives have the same sign,
the two parts of the sum compete with each other. (This leads to a qualitative
explanation for overtone suppression.) As $V_o$ increases, we may expect
the anharmonicity parameter $b$ to decrease (towards its Morse value). Assuming
that the dipole derivatives are approximately constant over the chosen $V_o$
range, the overtone intensity would increase with $V_o$ at a given $R$. The
plots also show the overtone is less suppressed at higher $V_o$.

\subsection{Effect of variation of the dipole function}
\label{sec:mecke}

\begin{figure}[!tbh]
	\includegraphics{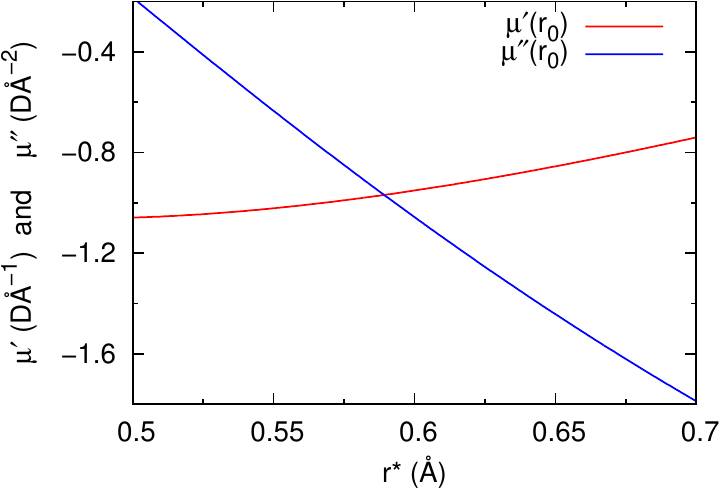}
	\caption{ Parameter sensitivity of the dipole derivatives.
	First and second derivatives of the Mecke function, $\mu_0(r) = \mu^* r
	\exp(-r/r^*),$ evaluated at $r=r_0=0.96$ {\AA} for a range of $r^*$ values.
	The OH bond $r^*$ value of $0.6$ \AA, given by Lawton and Child
\cite{Lawton:1980}, has been used in all earlier plots in this work. The first and second	 derivatives are relevant to the intensity of the fundamental and
overtone transitions, respectively. Note that the first derivative varies little for the
parameter
range shown whereas the second derivative varies by a factor of about five.
} 
	\label{fig:meckederv}
\end{figure}

The shape of $\mu_g(r)$ [Eq.~\eqref{eq:mugdef}] is dependent on that of the
diabatic dipole function, $\mu_0(r)$. We have used the two-parameter Mecke function
form for $\mu_0(r)$ in this work. As Eq.~\eqref{eq:mecke} shows, the parameter $r^*$
governs its
shape while $\mu^*$ gives it magnitude. We now analyse how the fundamental and
overtone intensities change when $r^*$ is varied around its Lawton-Child value
of 0.6 {\AA} for the OH bond. The results below are for symmetric H-bonds.

The value of $r^*$ marks the position of the Mecke function maximum. When
varied, it is useful to know how the first and second derivatives of $\mu_0$
change at the reference OH distance of $r=r_0=0.96$ \AA. Figure
\ref{fig:meckederv} shows that the first derivative 
$\mu'_0(r_0)$ changes within about 20\% as $r^*$
is varied from 0.5 to 0.7 \AA. However, the second derivative $\mu''_0(r_0)$
changes more substantially, doubling at 0.7 {\AA} and reducing at 0.5 {\AA} to 20\% of theoriginal value (at $r^*=0.6$ \AA). This suggests that changing $r^*$, and hence
$\mu_g$, might result in a noticeable but fractional change on the fundamental
intensity, but substantially alter the intensity of the overtone.

\begin{figure}[!tbh]
	\includegraphics{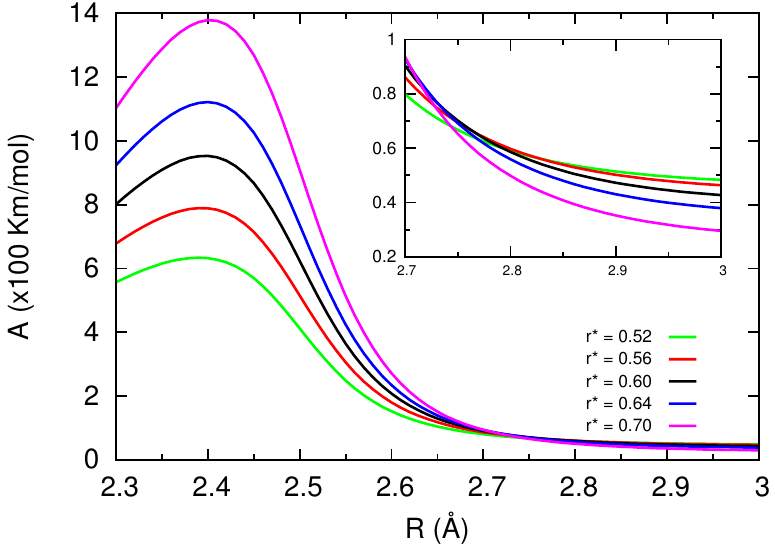}
	\includegraphics{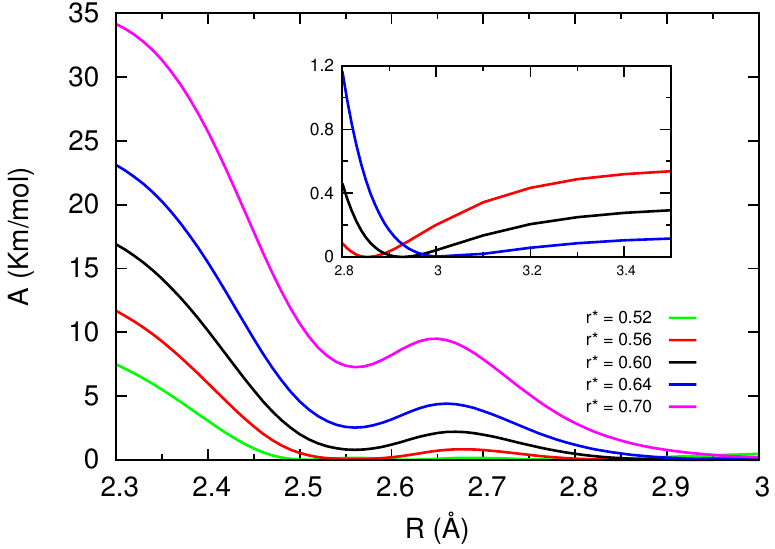}
	\caption{ 
	Variation of fundamental (top panel) and overtone (bottom panel) intensity
	with $R$ for different Mecke parameters $r^*$. The $r^*=0.6$ {\AA} curves
	are both the same as those in Figures \ref{fig:fund} and \ref{fig:Overtone}.
	The top panel shows that the fundamental enhancement is only somewhat
affected by $r^*$, especially for weak bonds, while the overtone intensities change more
dramatically.
	The fundamental intensities show a curious trend switch around 2.75 \AA,
	which is shown in the inset and analysed in the main text.
	}
	\label{fig:meckevar}
\end{figure}

Figure \ref{fig:meckevar} shows the $1^+ \leftarrow 0^-$ fundamental (top panel)
and $2^+ \leftarrow 0^-$ overtone (bottom panel) intensities as a function of
$R$ for different $r*$ values. For the fundamental, the intensity changes with
$r^*$ appear larger for strong H-bonds (lower $R$). However, these are only a
consequence of the uniform scale of the plot's $y$-axis. The intensities for
successive $r^*$ values generally differ by about 10-20\% at both large and
small $R$. In effect, variation in the fundamental intensities with the shape
parameter of the diabatic dipole function is modest. We note that for $R \gtrsim
2.8$ \AA, the intensity is lower for larger $r^*$, consistent with the Mecke
function derivatives discussed above. But this trend is reversed for $R \lesssim
2.75$ \AA: The intensity enhancement is larger (smaller) for larger (smaller)
$r^*$.  We briefly analyse this trend.

\newcommand{\avgS}{\langle S \rangle}

All components of the transition moment integral $\left\langle \phi_{1+} \mid
\mu_g \mid \phi_{0-} \right\rangle$ vary with $R$, but only $\mu_g(r)$ changes with
$r^*$. We rewrite the integral as $\mu_{g,n} \left\langle \phi_{1+} \mid \mu_g /
\mu_{g,n} \mid \phi_{0-} \right\rangle = \mu_{g,n} \avgS$. We take $\mu_{g,n}
\equiv \mu_g(r=r_{node})$, where $r_{node}$ is the (non-central) node of the
wavefunction product (shown in Figure \ref{fig:mu_condon}) for that $R$.  Note
that $r_{node}$ does not shift with $r^*$ at a given $R$, and therefore provides
a common reference point at that $R$. In this manner, the transition moment is
separated into a shape part, $\avgS$ and an overall magnitude, $\mu_{g,n}$. Though not
shown,
we found that plots of $\mu_g/\mu_{g,n}$ for various $R$ and $r^*$ look nearly
the same. Table \ref{tb:mecke} shows the intensity contributions of these pieces
for $r^*=0.56$ and 0.64 \AA, relative to those at $r^*=0.6$ \AA. For the shorter distances
($R=2.6$ and 2.4 \AA), the relative intensities ($A$ ratios) are about the same
as the relative $\mu_{g,n}^2$. The ratio of $|\avgS|^2$ is nearly unity, so the
shape of the dipole function plays a minor role. However, for weak H-bonds, the
shape appears to play a role. At $R=2.8$ {\AA}, it overrides the effect of
$\mu_{g,n}$. 

For the overtone, the bottom panel of Figure \ref{fig:meckevar} shows that although
overall shape remains about the same, the intensity drops strongly with
decreasing $r^*$.  This appears in agreement with the variation in
$\mu''_0(r_0)$ discussed at the start of this section. However, for weak
H-bonds, shown in the inset is a trend reversal. Indeed, in this region the
$\mu''_0$-based analysis is expected to be more valid. This suggests that
overtone intensities in this context are perhaps not easily analysed by way of
derivatives, and that details of the  transition moment integrand, viz.
$\psi_{2^+}\mu_g\psi_{0^-}$, do matter. Another aspect that the inset points to
is that the extent and range of overtone suppression in the weak H-bonding range
is a sensitive function of $r^*$.

\begin{table}[t]
	\caption{
	Contribution to the fundamental intensity at different value of Mecke
	parameter $r^*$. Compared are the total intensity ($A$), the scale factor
	($\mu_{g,n}$) and the matrix element of $\mu_g/\mu_{g,n}$, denoted $\avgS$.
	All results are reported as ratios relative to the results for $r^*=0.6$;
    the ${}^o$ superscript in the table header indicates the value of the
	quantity for $r^*=0.6$. The final column is obtained as a ratio of the 
	middle two.
	}
	\label{tb:mecke}
	\begin{tabular}{c@{\hspace{3mm}}c@{\hspace{3mm}}c@{\hspace{3mm}}c}
	\hline\hline
	$R$ (\AA)  & $A/A^o$ & $|\mu_{g,n}/\mu_{g,n}^o|^2$ & $|\avgS/\avgS^o|^2$\\ 
	\hline
	\\
	\multicolumn{4}{c}{$r^*=0.56$} \\ \hline
	2.8 &	 1.02 &  0.89  &    1.15 \\
	2.6 &	 0.87 &  0.89  &    0.97 \\
	2.4 &	 0.83 &  0.80  &    1.03 \\
	\\
	\multicolumn{4}{c}{$r^*=0.64$} \\ \hline
	2.8 &    0.95 &  1.11  &    0.86 \\
	2.6 &    1.13 &  1.10  &    1.02 \\
	2.4 &    1.12 &  1.18  &    0.94 \\ 
	\hline\hline
	\end{tabular}
\end{table}

\section{Comparison with previous work}
\label{sec:comparison}

We have already noted in earlier sections that that our results for the
fundamental enhancement in Figure~\ref{fig:fund} (solid line) and the
corresponding $A_H/A_D$ ratio in Figure~\ref{fig:ratio} with SGIE are in overall
agreement with experimental ranges summarised by Bratos {et al.}
\cite{Bratos:1991}; see Sections \ref{sec:FI} and \ref{sec:isotope}. We now
consider some specific molecular systems.

\subsection{Symmetric H-bonds}
\label{sec:comp_symm}

Bournay and Marechal \cite{Bournay:1973} measured the isotope intensity ratio
for acetic acid dimers in the gas phase (which have $R \simeq 2.68$ \AA
\cite{Derissen:1971}), finding a ratio of $2 \pm 0.2$ for the transition
probabilities (i.e., $|\mu_{fi}|^2$). Owing to a marked departure from the harmonic value
$\sqrt{2}$, they suggested the value to be anomalous, and attributed it to a
breakdown of the Born-Oppenheimer approximation.  
However, our model is within this
approximation. At that O-O distance, Fig.~\ref{fig:ratio} gives $A_H/A_D \approx
2.6$, while the frequency ratio $\nu_H/\nu_D \approx 1.3$ from Figure 8 of
Ref.~\onlinecite{McKenzie:2014}. This yields a transition probability of about
2.0, which agrees with their measurements. In contrast, our estimate of about 5
for the H isotope's $|\mu_{fi}|^2$ enhancement compared to the monomer is much
lower than their experimental estimate of about 32.

A number of theoretical and experimental studies have been performed on the
Zundel cation, H$_5$O$_2^+$, which has $R \simeq 2.5$ \AA.  More recently, Tan
and Kuo \cite{Tan:2015} studied (CH$_3$OH)$_2$H$^+$, which has $R \simeq 2.4$
\AA. For the Zundel cation, one of the peaks around  1000 cm$^{-1}$ is
identified with the proton transfer motion, which would correspond to the $0^-
\leftarrow 0^+$ transition in the notation of present work; see our footnote in
Ref.~\onlinecite{zundelnote}. Unfortunately, an experimental measurement of the
absolute intensity seems unavailable.  Theoretical studies also give the
relative intensities of this mode to be 3-10,\cite{Vener:2001} 8,
\cite{Vendrell:2007} and 20-40 \cite{Guasco:2011} times the intensity of the OH
stretches of the end groups.  For (CH$_3$OH)$_2$H$^+$, the OH stretch at 1010
cm$^{-1}$ (also $0^- \leftarrow 0^+$ in our notation) was computed to have an
intensity of 2567 km/mol \cite{Tan:2015}. 

In the present work for
 $R=2.5$ {\AA} and 2.4 {\AA}, the $0^- \leftarrow 0^+$ transition 
 has frequencies of about 164 and 750 cm$^{-1}$, respectively.
Clearly, these are lower than those of the aforementioned works. The
corresponding intensities are about 335 and 600 km/mol, or enhancements of about
9 and 15. (Note that these are relative to $1^+ \leftarrow 0^-$ at $R=6.0$ \AA.)
At the same $R$ values, $1^+ \leftarrow 0^-$ frequencies are 1620 and 1780
cm$^{-1}$, which err on the higher side compared to the Zundel and
(CH$_3$OH)$_2$H$^+$ cations.  The corresponding intensities are about 625 and
950 km/mol, i.e. 16 and 24-fold enhancement.

We end this section with a brief comparison with a particle in a box (PIB)
model. For short, strong H-bonds the potential appears similar to that for a
PIB of width $L=R-2r_0$ with quantum numbers $n=1,\,2,\,3,\,\ldots$
\cite{McKenzie:2014}. The corresponding transition dipole moment for a
transition $n_f \leftarrow n_i$, is only non-zero when  $n_f - n_i$ is odd, for
which $A_{if} \propto (n_f^2 n_i^2)/(n_f^2 - n_i^2)^3$ is box-length
independent. 
Here, the three transitions 
$3 \leftarrow 2$,
$4 \leftarrow 1$, and
$5 \leftarrow 2$, in the PIB.  
correspond to 
$1^+ \leftarrow 0^-$,
$1^- \leftarrow 0^+$, and
$2^+ \leftarrow 0^-$,
respectively, in the strong H-bond case. In the PIB, the three transitions have
the intensity ratios about 60:1:2.  Figures \ref{fig:fund}, \ref{fig:Overtone}
and \ref{fig:overt2} suggest that the the ratios are roughly comparable, but
are still clearly $R$-dependent unlike the PIB case.

\subsection{Asymmetric H-bonds}
\label{sec:comp_asym}
The molecular system discussed below are weak H-bonds.
For numerical comparisons, we will use our asymmetric model with $V_o=50$
kcal/mol.  Note, however, that this choice of $V_o$ is not special. Our results
do vary somewhat with $V_o$, as Figure \ref{fig:overtone_asymm} demonstrates.
For the fundamentals alone, we additionally quote our symmetric model results
for contrast. 
Also, if the H-bonded O-O distance was not directly available from the cited
work, it was estimated using the given OH fundamental red-shift and Figure
\ref{fig:omega} .

For the fundamental of ethanol dimers, Provencal {et al.} \cite{Provencal:2000}
calculated an enhancement of 10-20 relative to the monomer in the double harmonic
approximation. 
For intramolecularly H-bonded propane- and butanediol, Howard and Kjaergaard
\cite{Howard:2006} report the OH stretch intensity to be enhanced 4-11 times for
different conformers. These H-bonds have $R\simeq2.8-2.9$ {\AA}, for which our
enhancement factors are 1.3-1.5 for the symmetric model and 1.07-1.12 for the
asymmetric model.
Suhm and co-workers' \cite{Scharge:2008} experiments on 2,2,2-trifluoroethanol
dimers show an intensity enhancement of 4.0 $\pm$ 0.8 for the fundamental of the
donor O-H compared to the acceptor O-H.  Our values are about 1.26 and 1.04 for
the symmetric and asymmetric cases.
In general, the enhancement from our calculation for $R \gtrsim 2.7$ {\AA} 
is at most $\sim 2$ with the symmetric model and $\sim 1.2$ with the asymmetric
model (see inset of Figure~\ref{fig:fund}), both of which are smaller than
values in the literature.

For the same molecules, however, our overtone suppression estimate compares
somewhat more favourably.
Suhm and coworkers reported a value of $A/A_{free}$
 for 2,2,2-trifluoroethanol dimer to be
$0.3 \pm 0.1$ \cite{Scharge:2008}. Our estimate is 0.63.
Calculations by Howard and Kjaergaard \cite{Howard:2006} for propane- and
butanediols indicate a suppression from 0.43 to 0.15, with
lower values for butanediols. Our estimates are consistent with this relative
ordering of magnitudes and in the
range 0.41 to 0.20.
In general, literature values of $A/A_{free}$ for the overtone are about 0.5 to 0.1, the
smaller values pointing to stonger H-bonds. Figure \ref{fig:overtone_asymm}
indicates (for $R$ between 2.8 and 3.0 \AA) that our estimates are in about that
range, allowing for variation of the asymmetry parameter $V_o$.


Finally, we discuss another metric, namely the fundamental-to-overtone intensity
ratio, $A_1/A_2$. This ratio is typically about 10 for monomers, and is reported
to increase by over an order of magnitude with H-bonding \cite{Scharge:2008,
Kollipost:2014, Phillips:1998} in the weak region.
For 2,2,2-trifluoroethanol dimer, Scharge {et al.} \cite{Scharge:2008} report
$A_1/A_2=400\pm100$ and $30\pm10$ for the donor and acceptor O-H
bonds, respectively. Our $A_1/A_2$ ratios are
about 366 and 218, respectively. The experimental monomer ratio of 
$13\pm2$ is smaller than our ($R=6.0$ \AA) estimate of about 122.
A more recent work from the Suhm group on the dimers of methanol, ethanol and
t-butyl alcohol \cite{Kollipost:2014} gives the $A_1/A_2$ ratio for the donor
O-H as $320\pm90$, $400\pm100$ and $1000\pm400$. Using values of $R$ deduced
from redshifts, our ratios are $\approx$ 493, 583, and 711, in reasonable accord
with experiment. 

We also mention that some O-H$\cdots$Y-type asymmetric complexes have been
analysed, e.g.  F$^-\cdot$H$_2$O \cite{Yates:1988}  and Cl$^-\cdot$H$_2$O
\cite{McCoy:2014} in theoretical studies. The former has a strong H-bond, for
which an OH fundamental intensity enhancement of about 35 was computed (in the
double harmonic approximation). For the chloride complex, it was found to be 50
using an anharmonic treatment. It also showed overtone suppression of 0.35. We
have not attempted any numerical comparisons for these cases, since our model is
parametrized for O-H$\cdots$O systems.

\section{Summary and Concluding Remarks}

We have discussed the intensity variation of the
fundamental and overtone transitions in O-H$\cdots$O type H-bonds.  The results
are based on a diabatic two-state potential model and a Mecke form for the
diabatic dipole moment. These yield a ground adiabat and associated adiabatic
dipole moment along the H-atom transfer coordinate. The latter along with
one-dimensional vibrational wavefunctions were used to compute the intensities
for a range of O-O distances. Over this range, the H-bond varies from weak to
strong. Also analysed are the role of donor-acceptor asymmetry (i.e. difference
in their pK$_a$'s) as well as the effect of the shape of the 
Mecke function for the dipole moment.

For the OH fundamental, we find that the intensity is enhanced compared to the
free OH over all relevant O-O distances, ranging from a factor of under 2 for
weak H-bonds to about 20 for strong bonds. We show that the non-linearity of
the dipole moment is important, especially for medium and strong H-bonds, and
therefore the Condon approximation is not suitable. The H/D isotope effect was
analysed in terms of the fundamental intensity ratio, which is found to be
non-monotonic with H-bond strength. A maximum occurs for this ratio at
the donor-acceptor distance 
$R$ of about 2.5 {\AA}, and the secondary geometric isotope effect
plays an important role in the height and position of this maximum.
 For the OH overtone, our
model finds intensity suppression for weak H-bonds, and shows variability in
magnitude and $R_{\text{OO}}$ range depending on whether we consider symmetric
or asymmetric bonds. For medium and strong H-bonds, enhancements in the
intensity are seen with the symmetric model, going up to 50 times the free OH
value. This new finding suggests that overtones should be easily experimentally
visible for such H-bonds.

Our results are generally consistent in trends but differ in numbers with
previous work, including both experimental and
theoretical studies. In particular, our enhancements in fundamental intensities
for weak H-bonds are clearly lower. Comparisons of overtone suppression in the
same region with the asymmetric model fare somewhat better.  
These comparisons suggest that in regime of weak asymmetric
bonds that our simple model may be missing some key physical ingredient.
Variations in shape of the dipole moment function lead to modest fractional change in the
intensity of the fundamental, but to larger changes for the overtone.

Studies of H-bond intensities offer an excellent point of comparison for experiment
and theory, owing to the large spread of bonding strengths and topologies. In
the present context of O-H$\cdots$O H-bonds, with a few exceptions such as
H$_5$O$_2^+$ and (CH$_3$OH)$_2$H$^+$, most detailed studies have mainly focussed
on specific systems in the weak H-bonding regime. Experiments on
symmetric medium and strong H-bonded systems  are
desirable. Some possible candidates are carboxylic acid dimers ($R \simeq 2.45$
\AA), HCrO$_2$ ($R \simeq 2.49$ \AA), porphycenes\cite{Ciacka:2015},
and proton sponges\cite{Horbatenko:2011}, for which the
fundamental, first overtone, and isotope effect could be measured and analysed. 
Slightly asymmetric biomolecular systems with strong H-bonds
that could be investigated include mutated GFP\cite{Oltrogge:2015}, 
photoactive yellow protein\cite{Nadal:2014} and the enzyme KSI\cite{Wang:2014}.

\acknowledgments
We thank Seth Olsen for helpful discussions.

\bibliography{infrared} 

\begin{thebibliography}{55}
\bibitem[{zun()}]{zundelnote}
\bibinfo{note}{{There is debate about the correct identification of features in
  the IR spectra, due to large anharmonic and intermode coupling effects,
  requiring a high-dimensional potential energy surface.}}
\end{thebibliography}
\end{document}